\documentclass[11pt]{article}

\usepackage[margin=1.1in]{geometry}
\usepackage{amsmath,amssymb,mathtools}
\usepackage{graphicx}
\usepackage{booktabs,tabularx}
\usepackage{microtype}
\usepackage[font=small,labelfont=bf]{caption}
\usepackage{xcolor}
\usepackage{cite}
\usepackage[colorlinks=true,linkcolor=blue!50!black,citecolor=blue!50!black,urlcolor=blue!50!black]{hyperref}
\usepackage[capitalise,noabbrev]{cleveref}

\title{Junction Hook Errors in the Color Code}
\author{Salahedeen Issa \qquad Margaret Martonosi\\[2pt]
  \small Department of Computer Science, Princeton University\\
  \small \href{mailto:salahedeen@princeton.edu}{\texttt{salahedeen@princeton.edu}}}
\date{September 2026}

\begin{document}
\maketitle

\begin{abstract}

Color-dependent syndrome extraction schedules have been proposed to mitigate hook errors in the triangular color code. Recent work by Kishony and Fowler~\cite{KF} identifies distance-reducing hook errors by the type of anyon for which they provide a shortcut and chooses the CNOT ordering for stabilizers of each color to avoid these errors. We identify an additional class of distance-reducing hooks, which we call junction hook errors, that is not captured by this classification. The two
data qubits of a junction hook error terminate anyon strings of two different colors at the point
where the three colored strings of a logical operator meet, which we call the junction. As such, the hook error does not correspond to a
single type of anyon, yet it still reduces the distance. For distances $d$=7,9,11, we verify that every non-corner plaquette admits a scheduled hook error that, together with $d-2$ single data qubit faults, produces an undetectable logical error using $d-1$ faults. Each of these is a junction
hook error. Because of this, every stabilizer, including those in the bulk, still requires additional protection from hook errors (such as flag qubits~\cite{CR,CB}) to restore the full circuit-level distance with this schedule. A minimum weight
logical operator has at most one junction, however, so junction hook errors cannot accumulate on one minimum weight
logical operator in the way that single-color malign hook errors can. This helps explain why the schedule of Ref.~\cite{KF} still improves the circuit-level distance to \mbox{$d-\lfloor (d+3)/6\rfloor$}, rather than the $d/2$ distance associated with spatially uniform color code scheduling. 

\end{abstract}

\begin{figure}[t]
  \centering
  \includegraphics[width=0.74\linewidth]{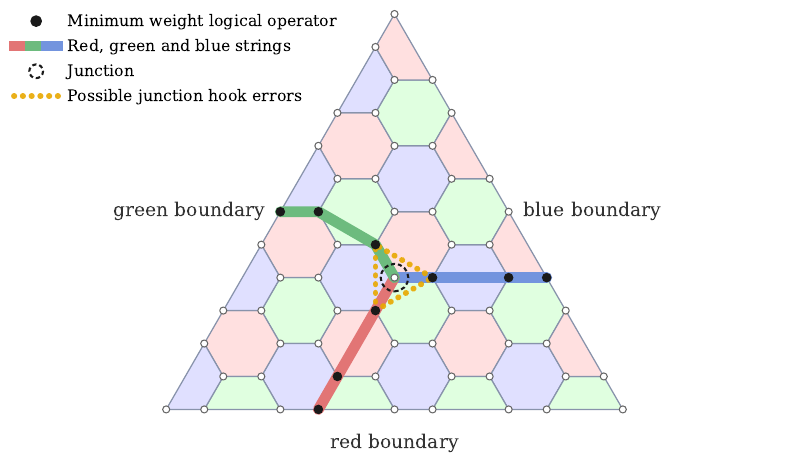}
  \caption{An example of a junction hook error on the $d=9$ triangular color code. The black dots form a minimum
  weight logical operator, which consists of a red, a green and a blue string. Each string begins
  on the boundary of its own color, and the three strings meet at the junction (dashed circle). The junction is a
  data qubit that does not belong to the logical operator but whose three neighbors do. These strings are all either perpendicular or ``nearly'' perpendicular to the boundary of the same color. The
  junction lies on one stabilizer of each color, and has two neighbors on each stabilizer that both lie on the minimum weight logical operator. A hook error on these two data qubits is what we call a junction hook error (dotted yellow). On the red stabilizer, for example, one of the two
  data qubits ends the green string and the other ends the blue string. The hook error provides a
  shortcut for strings of two different colors rather than for a single type of anyon. Together
  with the remaining seven data qubit errors, any one of these hook errors produces an undetectable
  logical error with 8 faults, one fewer than the code distance. Under the schedule of Ref.~\cite{KF}
  (\cref{fig:kf}(b)), all three pairs are weight-2 hook errors, at positions 5 and 6 of the red
  stabilizer, 1 and 2 of the green stabilizer, and 5 and 6 of the blue stabilizer.}
  \label{fig:junction}
\end{figure}

\section{Introduction}

A minimum weight logical operator in the triangular color code can be represented as strings of different colors propagating from the boundaries and meeting somewhere in the patch. Each colored string originates from the boundary with the same color, and crucially, only propagates in a manner that is either perpendicular or "almost" perpendicular to that boundary.

\begin{figure}[t]
  \centering
  \includegraphics[width=0.94\linewidth]{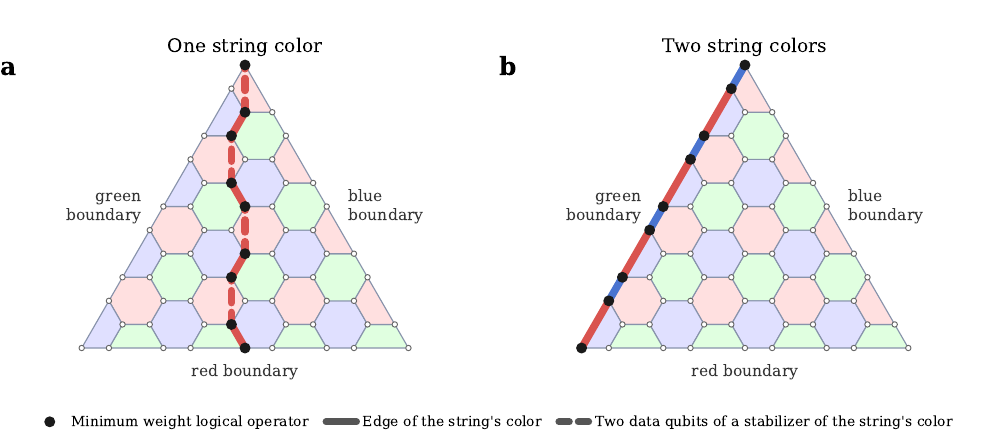}
  \caption{Minimum weight logical operators do not necessarily involve strings of all three colors, as shown here on the $d=9$ triangular color code. A string advances in one of two ways: along an edge of its own color
  (solid), or across a stabilizer of its own color, through two of that stabilizer's data qubits
  (dashed). Both are oriented perpendicular or nearly perpendicular to the boundary of the
  string's color. (a)~A red string that runs from the red boundary into the top corner, where the
  green and blue boundaries meet. It alternates between red edges and pairs of data qubits on red
  stabilizers, and ends on the corner data qubit, which belongs only to the red corner stabilizer,
  so the green and blue strings have length zero. (b)~The logical operator along the green
  boundary, where the green string has length zero and the logical operator alternates between
  red and blue edges. The blue edges are sides of the red stabilizers along this boundary. The red edges are
  the long sides of the weight-4 blue stabilizers. In both of these cases, there can be no junction hook error, as there is no junction. }
  \label{fig:strings}
\end{figure}

The arXiv preprint ``Color code off-the-hook'' by Kishony and Fowler~\cite{KF} develops a new
syndrome extraction schedule for the color code based on this. They identify that, in a minimum weight logical operator, strings of a given color only propagate through the code in two ways: on an edge of that color, or on two data qubits in a stabilizer of that color whose line is almost perpendicular to the boundary of that color (\cref{fig:strings}). Any hook error that provides a shortcut for a string of a given color is said to be malign, and the schedule is designed to avoid them. Since the
strings of each color traverse the code in different directions, different hook errors are
malign on stabilizers of different colors. Consequently, the authors give each color of stabilizer its own
CNOT ordering, chosen so that each color's hook errors do not provide shortcuts for these strings. The
authors report that the resulting circuit improves the
circuit-level distance from $d/2$ to
\mbox{$d-\lfloor (d+3)/6\rfloor$}, and attribute this improvement to the fact that the schedule ``avoids all malign hook errors in the bulk,'' while ``at the boundary, no single hook error alone reduces the distance''~\cite{KF}. 

Identifying malign hook errors by their color, however, does not account for all malign hook errors in the code. When a logical operator has strings of all three colors, there is a
point in the bulk at which they meet.
We call this point the junction of the logical operator (\cref{fig:junction}). A weight-2 hook error
whose two data qubits are the ends of two of the strings, on either side of the junction, provides a
shortcut for two strings of different colors at the same time. While such a hook error does not correspond to a single color, it still acts as a shortcut by connecting different colored strings at the junction. We call these \emph{junction hook errors}. The schedule of Ref.~\cite{KF} places a distance-reducing junction hook error on every non-corner stabilizer, so that a single hook error, together with $d-2$ data qubit errors, can produce an undetectable logical error. Junction hook errors
therefore provide counterexamples to the statement in Ref.~\cite{KF} that the schedule ``does not admit any single hook error that alone reduces the circuit-level distance, except at the corners of the system.''

A minimum
weight logical operator has at most one junction where all the strings meet, so junction hook errors cannot accumulate on a minimum weight logical operator in the way that single-color malign
hook errors can. This is why the schedule still manages to improve the distance greatly by avoiding single-color malign hook errors (which can accumulate on the same logical operator).

\section{Background and Definitions}

\subsection{Color Code Logical Operators}

A logical operator in the 2D triangular color code~\cite{BM} is any odd-weight chain of data qubit operators that
commutes with every stabilizer in the code, meaning that it overlaps each stabilizer on an even
number of data qubits. Such a chain necessarily connects all three boundaries~\cite{KMTEBB}. A minimum weight
logical operator is any weight-$d$ chain of data qubits that satisfies the same conditions.
Since the color code is self-dual, we only consider $X$-type operators below, but the same
arguments apply to $Z$-type operators.

\begin{figure}[t]
  \centering
  \includegraphics[width=\linewidth]{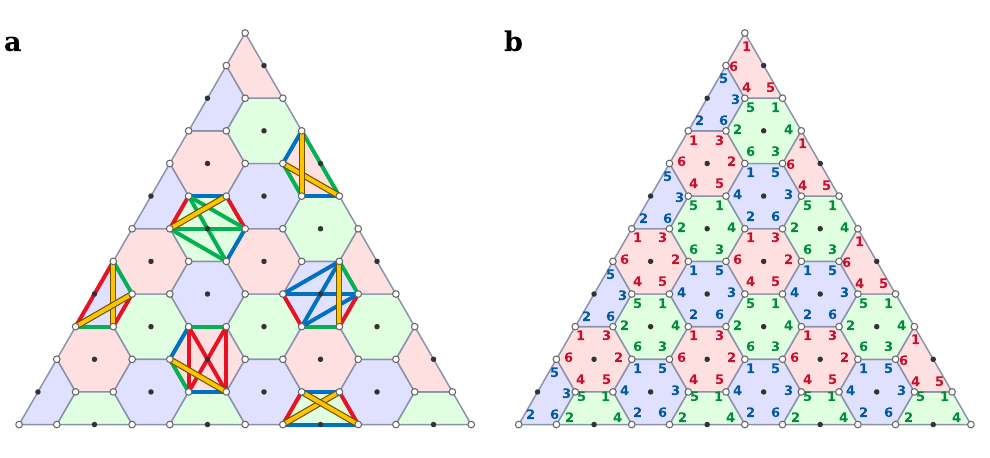}
  \caption{Malign hook errors and the schedule of Ref.~\cite{KF}, drawn on the $d=9$ triangular
  color code in the layout of Fig.~1 of Ref.~\cite{KF}. (a)~Fig.~1(a) of Ref.~\cite{KF}, with
  the junction hook errors added in yellow. As in Ref.~\cite{KF}, the red, green and blue pairs
  are the single-color malign pairs, shown for one bulk hexagonal stabilizer and one weight-4
  boundary stabilizer of each color, and colored according to the type of anyon for which they
  would provide a shortcut. On each hexagonal stabilizer, the yellow pair joins the ends of the
  two strings whose colors differ from that of the stabilizer. On the weight-4 stabilizers, both
  diagonals are shown in yellow. They differ by the stabilizer itself and so describe the same
  hook error. One of them is a junction pair, and the other is perpendicular to the boundary of
  the stabilizer's color (\cref{sec:counterexamples}). (b)~The schedule. Each number is the time step
  (1--6) at which that data qubit is coupled to the stabilizer's ancilla qubit. The schedule is
  the one provided with the authors' circuits~\cite{repo} ($R=[1,3,2,5,4,6]$,
  $G=[5,1,4,3,6,2]$, $B=[1,5,3,6,2,4]$, listed for the top-left, top-right, right, bottom-right,
  bottom-left and left data qubits), and it matches Fig.~1(b) of Ref.~\cite{KF}. The weight-4
  boundary stabilizers use the same ordering, with the two missing data qubits omitted.}
  \label{fig:kf}
\end{figure}

\subsection{Hook Errors}

In the circuits of Ref.~\cite{KF}, each stabilizer is measured using a single ancilla qubit,
which is coupled to the data qubits of the stabilizer by CNOTs in the order given by the
schedule (\cref{fig:kf}(b)). An error on the ancilla qubit after the $k$-th CNOT propagates onto every data
qubit coupled after it. Such correlated errors, known as hook errors, are a central consideration in the
design of syndrome extraction circuits~\cite{DKLP,TS,KFsurface}, and the choice of CNOT ordering for
the color code has been studied in several works~\cite{BKS,ccs,GJ}. For a weight-6 stabilizer, an error after the fourth CNOT propagates
onto the data qubits at positions $(5,6)$ of the schedule, and an error after the second CNOT
propagates onto positions $(3,4,5,6)$, which is equivalent to positions $(1,2)$ up to the
stabilizer itself. These are the two weight-2 hook errors of the stabilizer. The weight-4
boundary stabilizers use the same schedule with the two missing data qubits omitted, so their
weight-2 hook error acts on the last two data qubits in their schedule (or equivalently, the
first two).

\subsection{Anyons in the Color Code}
In the anyon picture of the color code~\cite{KMTEBB}, an error on a data qubit flips the
stabilizers of the opposite type that contain it (three in the bulk, fewer on the boundary), creating an anyon
on each of these plaquettes. Each anyon
takes the color of its plaquette. Each edge of the lattice carries the color of the two plaquettes
that it connects at its endpoints, and errors on both ends of an edge of color $c$ move an anyon
of color $c$ from one of these plaquettes to the other. A logical operator ``can
be viewed as propagating a red anyon from the red boundary, a green anyon from the green boundary,
and a blue anyon from the blue boundary to the same point in the bulk, where all three annihilate
together''~\cite{KF}. We refer to the three paths as the red, green and blue strings of the
logical operator, and to the point where they meet as its junction. In terms of stabilizers, an anyon of color $c$ is a stabilizer of color $c$ whose measurement
outcome has been flipped.

In a minimum weight logical operator, a string of color $c$ advances either along an edge of
color $c$, or across a stabilizer of color $c$ through two of its data qubits, in both cases in a
direction perpendicular or nearly perpendicular to the boundary of color $c$ (\cref{fig:strings}).
When the strings meet at a corner or along a boundary, one or two of them have length zero, so a
minimum weight logical operator can also be made of strings of only one or two colors.

\subsection{Malign Hook Errors}\label{sec:malign}
A malign hook error, according to Ref.~\cite{KF}, is any hook error that alone reduces the distance
of the code. It does so by providing a shortcut for a logical operator to cross through. In
other words, if a hook error can be scheduled to propagate onto two data qubits that lie on the
same minimum weight logical operator, that hook error is considered to be malign. With this definition, junction hook errors should also be considered malign.

\subsection{Single-Color Hook Errors}\label{sec:definition}
Ref.~\cite{KF} identifies malign hook errors as shortcuts along a single string. For correlated errors that lie on an edge, ``shortcuts for logical operators are
enabled by correlated errors on edges of each color that are oriented nearly perpendicular to the
boundary of the same color,'' and a pair that does not lie on an edge ``is malign when both
outgoing half-edges, which share the color of the plaquette hosting the error, are oriented nearly
perpendicular to the boundary of that color''~\cite{KF}. In both cases the pair is assigned the
color of the string that it would shorten, and we refer to these as single-color malign hook
errors.

\section{Junction Hook Errors}\label{sec:junction}
The single-color malign hook errors, however, do not incorporate every malign hook error. Consider a
minimum weight logical operator $L$ with a junction at a data qubit $v \notin L$ whose
three neighbors all belong to $L$, so that each of the three strings of $L$ ends on one neighbor of $v$. We call a weight-2 hook error a
\emph{junction hook error} if its two data qubits, which we call a \emph{junction pair}, are the
two neighbors of $v$ on one of the stabilizers that contain $v$. Such a hook error is contained in $L$, so together with the remaining
$d-2$ data qubits of $L$ it produces $L$ using $d-1$ faults. On a stabilizer of color $c$, its two
data qubits terminate the two strings whose colors differ from $c$. A junction hook error therefore provides a
shortcut for two strings of different colors at once, and is not described by a single type of
anyon.

\subsection{Junction Hook Errors in the Schedule}\label{sec:counterexamples}

\begin{figure}[p]
  \centering
  \includegraphics[width=\linewidth]{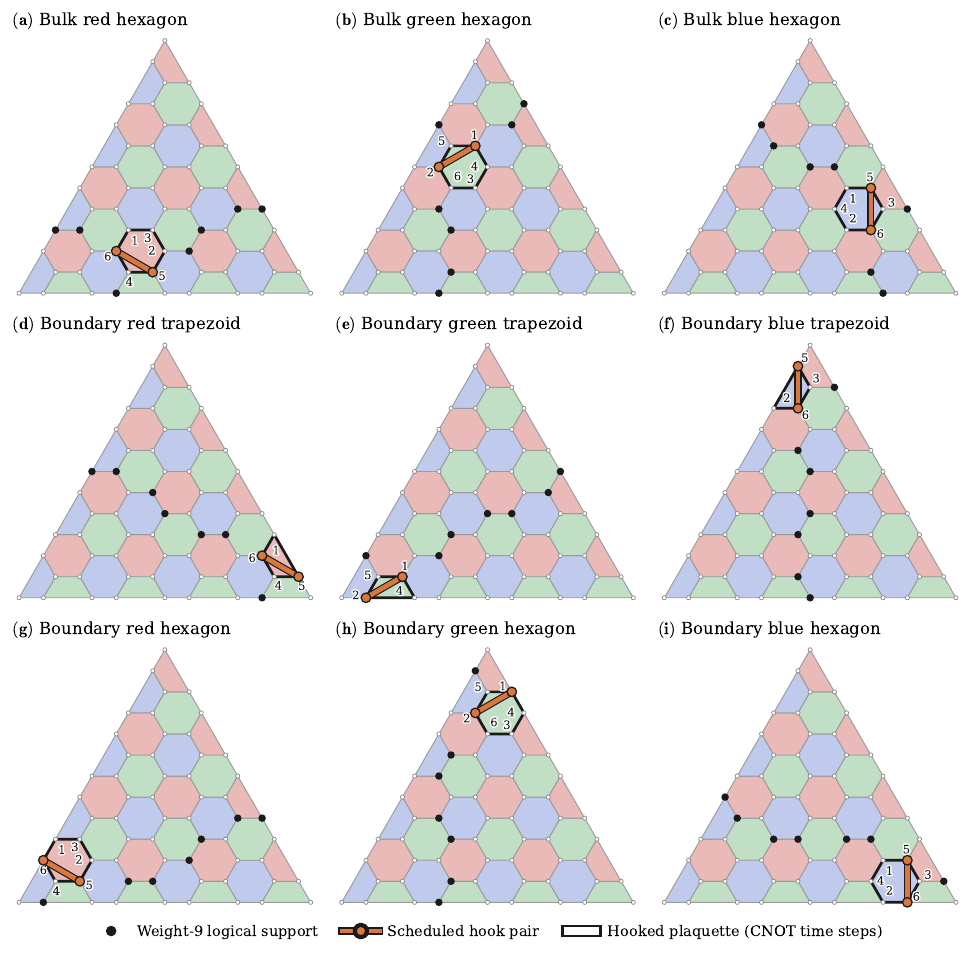}
  \caption{A junction hook error on every type of non-corner stabilizer under the schedule of
  Ref.~\cite{KF}, shown here at $d=9$: (a--c) bulk stabilizers, (d--f) weight-4 boundary
  stabilizers, and (g--i) weight-6 boundary stabilizers. In each panel, the numbers beside the
  corners of the outlined stabilizer give the CNOT ordering from the schedule. A weight-2 hook error occurs on the
  data qubits at positions $(1,2)$ or $(5,6)$ of a weight-6 stabilizer, and on the last two data
  qubits of a weight-4 stabilizer, which is equivalent up to the stabilizer to the first two;
  panel (e) shows the first two, which form the junction pair. Each panel shows a minimum weight
  logical operator containing the hook error, which in every case is a junction hook error
  (\cref{fig:junction}). The hook error and the remaining seven data qubit errors
  make up an undetectable logical error with 8 faults, one fewer than $d=9$.}
  \label{fig:counterexamples}
\end{figure}

With these definitions in mind, we now look at the individual hook errors produced by the
schedule of Ref.~\cite{KF}, both in the bulk and on the boundary stabilizers. On every red stabilizer in the bulk, the pair formed by the lower-right and middle-left data
qubits, which the schedule of Ref.~\cite{KF} couples at time steps 5 and 6 (\cref{fig:kf}(b)),
is a junction pair (\cref{fig:junction}). In the minimum
weight logical operator of \cref{fig:junction}, the lower-right data qubit ends the blue string and the
middle-left data qubit ends the green string. Since it does not shorten a single string, it is not identified as malign in Ref.~\cite{KF}. The same holds for the corresponding pairs on the green and blue stabilizers (positions 1
and 2, and positions 5 and 6, respectively). \Cref{fig:kf}(a) reproduces Fig.~1(a) of
Ref.~\cite{KF} with these junction pairs added. \Cref{fig:counterexamples} shows one
such hook error for each color of stabilizer in the bulk, on the weight-4 boundary stabilizers, and
on the weight-6 boundary stabilizers, together with a minimum weight logical operator that contains
it.

We checked every non-corner stabilizer of the code at $d=7$, $9$ and $11$. On each weight-6 stabilizer, exactly one of the two weight-2 hook
errors is contained in a minimum weight logical operator, and it is a junction hook error. The
weight-2 hook error of each weight-4 stabilizer is also a junction hook error. We also tested each of these hook errors in
the authors' circuits~\cite{repo} (simulated with
Stim~\cite{stim}), and found that an $X$ error on the ancilla qubit after the relevant CNOT, together with $X$
errors on the other $d-2$ data qubits of a minimum weight logical operator that contains the hook
error, fires no detector and flips the logical observable.

On the weight-4 boundary stabilizers, the hook error also has a second description. Its two
diagonals describe the same hook error up to the stabilizer, and while one of them is a junction
pair, the other is perpendicular to the boundary of the stabilizer's color. Multiplying a logical
operator that contains the junction pair by the stabilizer gives a minimum weight logical operator
that contains the other diagonal and has no junction next to it, and this diagonal satisfies the
criterion of Ref.~\cite{KF} for a single-color malign pair (\cref{sec:definition}). In fact, each of the three ways of pairing the four data qubits of a weight-4
stabilizer lies on a minimum weight logical operator, so no CNOT ordering gives these stabilizers
a hook error that leaves the distance intact. On the hexagonal stabilizers, by contrast, a hook
error lies on a minimum weight logical operator only at its junction.

\subsection{Circuit-Level Distance with Ancilla Noise Only in the Bulk}\label{sec:bulk-only}

Ref.~\cite{KF} states that the schedule ``avoids all malign hook errors in the bulk, thereby
preserving the full circuit-level distance.'' If the bulk stabilizers were free of malign hook errors,
fully protecting just the boundary stabilizers from errors would restore the circuit-level distance
to $d$. To check this, we removed all noise from the authors' circuit except for two-qubit
depolarizing noise after the CNOTs of the bulk stabilizers and single-qubit depolarizing noise
on the data qubits once per round. We then computed the exact circuit-level distance using the
integer program provided with their code~\cite{repo}. We find a circuit-level distance of 6 at
$d=7$ and 8 at $d=9$, one less than the code distance in both cases. With the noise on the bulk
CNOTs also removed, the distance is $d$, as expected.

\section{Implications}\label{sec:ramifications}

Practically, these results are mostly pertinent to constructions that combine the schedule with additional protection, such as flag qubits~\cite{CR,CB}, on the boundary stabilizers only, as suggested in Ref.~\cite{KF}. Since every stabilizer admits a distance-reducing hook error (on the corner stabilizers, all hook errors are malign~\cite{KF}), protecting only the boundary
stabilizers does not restore the full circuit-level distance, as shown in \cref{sec:bulk-only}.
Restoring the distance with this schedule would therefore also require
protecting the bulk stabilizers.

A single-color malign hook pair shortens one string,
and a string of a given color can pass through many plaquettes of that color. If the same pair were
a hook error on every plaquette of that color, a single logical operator could therefore use several
of them as shortcuts, and Ref.~\cite{KF} notes that most of these pairs, ``if replicated across all
plaquettes of the same color, would reduce the distance by a factor of two.''  A junction hook
error on a hexagonal stabilizer, in contrast, only shortens a minimum weight logical operator at its
junction, and a minimum weight logical operator has at most one junction. The hook error of a weight-4 stabilizer can also shorten a minimum weight
logical operator that has no junction next to it (\cref{sec:counterexamples}). Nonetheless, in the patches examined, no minimum weight logical operator contains two disjoint weight-2 hook errors
from the schedule, including those on the corner stabilizers. This agrees with the
circuit-level distance of $d-1$ found in \cref{sec:bulk-only}, and with the circuit-level distance
reported in Ref.~\cite{KF}, which the junction hook errors leave unchanged. For $d\geq 9$, the
remaining reduction, from $d-1$ to \mbox{$d-\lfloor (d+3)/6\rfloor$}, comes from combinations of
hook errors near the boundary, which Ref.~\cite{KF} calls fractional hook errors.

\section*{Acknowledgments}

We thank G.~Kishony and A.~Fowler for making their circuits and code publicly available, which
made the analysis in this note possible.

\section*{Code Availability}

The scripts used for the checks in this note (including the enumeration of logical operators,
the integer programs, the fault injection into the authors' circuits, and the figures) can be found here: {\nolinkurl{https://github.com/SalahedeenIssa/junction_hook_errors}}

\end{document}